**Phase mask pinholes as spatial filters for laser interference lithography**

G. Capraro,[1,2] M. Lipkin,[1] M. Möller,[1] J. Bolten[1], M. C. Lemme[1,2,*]

[1] Chair of Electronic Devices, Faculty of Electrical Engineering and Information Technology, RWTH Aachen University, Otto-Blumenthal-Straße 2, 52074 Aachen

[2] AMO GmbH, Otto-Blumenthal-Straße 25, 52074 Aachen

*lemme@amo.de

**Abstract:** Laser resonators have outputs with Gaussian spatial beam profiles. In laser interference lithography (LIL), using such Gaussian shaped beams leads to an inhomogeneous exposure of the substrate. As a result, dimensions of lithography defined features vary significantly across the substrate. In most LIL setups, pinholes are used as filters to remove optical noise. Following a concept proposed by Hariharan et. al. a phase mask can be added to these pinholes. In theory, this modification results in a more uniform beam profile, and, if applied as spatial filters in LIL, in improved exposure and hence feature size uniformity. Here, we report on the first successful fabrication of such elements and demonstrate their use in an LIL setup to reduce feature dimension variations from 47.2 % to 27.5 % using standard and modified pinholes, respectively.

1. **Introduction**

Laser Interference Lithography (LIL) [1] is a technique that enables the highly precise definition of periodic structures by recording the interference patterns of two laser beams in a photoresist sensitive to the laser's wavelength. The laser beams are typically Gaussian, so that the exposure dose in the resist varies with a maximum at the center of the interference pattern and a fall-off in both x and y directions. As a result, feature dimensions in the exposed resist also vary from the center towards the edges of the exposed substrate.

The Gaussian nature of a laser can be explained by considering the situation in a laser resonator, where only transverse Gaussian modes exist. The typical oscillating mode is the fundamental $TEM_{00}$, assuming that a laser beam travels almost parallel to the mirrors inside the resonator. Thus, the intensity of a laser beam extracted from the resonator can be modeled with a Gaussian function:

$$I(r, z) \propto I_0 \left(\frac{w_0(z)}{w(z)}\right)^2 \exp\left(\frac{-2r^2}{w^2(z)}\right) \qquad (1)$$

where r is the radial distance from the center axis of the beam, z is the axial distance from the beam's focus, also referred to as beam waist, w(z) is the radius at which the intensity is reduced to $1/e^2$ of its axial value and $w_0 = w(0)$ is the waist radius. As a consequence, laser sources from solid-state to gas or diode lasers have non-uniform spatial beam profiles.

Such non-uniform beam shapes are often undesirable, as a uniform illumination over an extended area is preferable in most applications [2]. For example, industrial applications like cutting and drilling require a beam with constant intensity to assure uniform physical interaction with the material. In display technologies, the entire region of interest needs to be illuminated uniformly. The same holds true for microfabrication techniques like holography and lithography. Here, a constant profile guarantees high pattern uniformity when fabricating electronic devices or optical elements. Therefore, the ability to tailor laser beam profiles both inside and outside laser resonators is a topic of interest [3], as it ensures optimum laser performance in a wide range of application scenarios.

Several concepts to modify the shape of laser beams have been proposed. Optical elements employed for beam shaping include Fabry-Perot etalons [4], sets of two aspheric transmissive phase plates [5] or plano-convex films fabricated using a non-uniform coating technique [6], that can convert a Gaussian beam into a super-Gaussian, a higher order distribution with a flat intensity profile near its maximum. Other methods make use of two positive convex aspheric lenses [7] or a set of holographic filters [8] to change the beam shape into a function known as a flat top profile (FTP), that has a flat part near its maximum. In addition, gratings with periodic features can be used to form an FTP by utilizing the diffraction properties [9].

All these methods suffer from one or more inherent drawbacks, although they provide beam profiles with a central flat portion: i) a severe reduction of the available laser intensity; ii) complex fabrication and need of customization for a certain wavelength and/or a specific use case; iii) increased alignment complexity and degradation of the beam spot shape, with the presence of speckles, due to additional optical elements added to a setup and/or iv) full width half maximum (FWHM) as a measure for laser intensity homogeneity only moderately increased and insufficient for many applications. Despite these drawbacks, those beam shaping methods have been successfully implemented in laser systems, where they provide suitable beam shapes for specific applications. Nevertheless, they do not adapt well in complex setups [10], where the beam needs to be expanded, usually by a spatial filter formed by an objective and a pinhole [11].

Grating waveguide structures (GWS) are an application where uniform feature dimensions are particularly important [12] [13]. These optical elements are a combination of planar waveguides and sub-wavelength gratings. They are formed by several layers acting as a highly reflecting mirror and a layer of line features with a period smaller than their operational wavelength. GWS are known to be a powerful solution to control key properties of light and in particular for tailoring the polarization and for realizing spectral and spatial beam shaping in high power laser systems [14] [15]. The uniformity of the two key properties of GWS, period and linewidth (LW), is essential to achieve high optical efficiency [16]. Here, we come back to LIL, because it is well-suited for GWS fabrication due to its high period precision, if the dimensional variations caused by the Gaussian beam profile can be countered, especially for large area GWSs. However, the methods for beam shaping previously described [1-8] can only be employed with limited efficiency when taking into consideration the need for uniform GWS features. Several of the described drawbacks apply, since a LIL setup is mounted over an extended area, contains several optical elements, and is used to expose wafer size samples.

Here, we experimentally demonstrate a concept for beam shaping based on modifying an existing circular pinhole as conventionally contained in spatial filters. The idea was first suggested by Hariharan et al. [10] as a method to modify a Gaussian beam. The principle concept is to add a ring-shaped transparent phase mask of a certain diameter to a standard circular pinhole. This element modifies a given Gaussian beam shape, leading to a significantly increased FWHM intensity. Moreover, the phase mask contributes to eliminating optical noise, such as speckles. It can be therefore used in complex systems where a constant beam with speckles-free constant beam shape is crucial, such as in holography or interferometry systems. The concept does not require additional optical elements and is ideally suited for LIL, where it is particularly important to prevent scattering from dust and reduce intensity losses. Moreover, the FWHM intensity of the beam is increased compared to the other beam shaping methods discussed above, and the flat part of the beam profile offers sufficient energy for highly precise LIL exposures. In this work, we have studied design parameters, used them to fabricate such "Hariharan pinholes" (HPH) and demonstrate their application to improve the quality of LIL gratings.

## 2. LIL setup

The custom-made LIL setup used in our experiment is shown schematically in Fig. 1. It consists of a resonator, pumped with a continuous-wave (CW) laser with a wavelength of $\lambda_{pump}$ = 532 nm, and is equipped with a non-linear birefringent crystal for frequency doubling. This provides a laser beam with a wavelength of $\lambda_{exp}$ = 266 nm with linear polarization. Outside the resonator, the beam passes first through an electronic shutter, used to switch on and off the laser beam. Then, a beam splitter divides it into two laser paths. They contain a $\lambda/2$ plate and a polarizer to be able to control the power in each arm. Finally, mirrors direct each beam to a spatial filter, making them speckles-free. Each of those spatial filters is formed by an objective to focus the beam into the pinhole, and a circular pinhole with a diameter of 2.1 µm that expands the beam.

The substrates to be exposed, in this case wafers, are positioned on a chuck where they are held in place by vacuum. The two expanded beams are steered such that the center of their interference patterns falls in the center of the substrates, where it is recorded in a resist sensitive for the exposure wavelength. A camera facing the chuck is used to create a grey scale image of the interference signal in order to monitor the beam shape and allow for a precise alignment. In reality, the beam is slightly ellipsoidal, with the long axis parallel to the x axis. This is due to the fact that, inevitably, the Gaussian beam hits the fluorescent screen under an angle.

Noise from the ambient and long-term drifts of the optical components cause an optical path difference between the two arms, and a phase difference between the two beams. The phase difference is controlled by a fringe locking system that makes use of a Mach-Zehnder interferometer, two photodiodes and a proportional–integral–derivative controller (PID circuit) connected to one arm of the setup, containing a Pockels cell. Thus, the interference pattern in the photoresist is stabilized and the contrast between exposed and non-exposed areas is maximized. The setup requires three signals to stabilize the frequency of the laser beam, by adjusting either the laser path length or the cavity length: the error signal from the Mach-Zehnder as input for the digital PID controller (AIXscan from AMOtronics ([17]), the modes inside the resonator and the error signal from the resonator corrective analog PID circuit.

3. Theory

The intensity of interference maxima of two Gaussian beams varies from a maximum at the center of the pattern and radially decreases following the Gaussian shape of those two beams. If used to expose a resist, as in LIL, this leads to periodic resist structures that have different dimension at the center and at the fringes of the interference pattern. This is enhanced in LIL, where the waves impinge with normal incidence on a circular pinhole of radius a, where they are diffracted, leading to inhomogeneous interference patterns. The relevant theory of diffraction is summarized in the supporting information.

Although the homogeneous width of a diffraction figure can be increased by reducing the size a of the pinhole, this approach quickly leads to excessive exposure times since the central intensity of the diffracted beam scales with $a^2$. Long exposure times lead to drift, compromising the quality of the exposed structures. We therefore explore another method of flattening the central peak of the diffraction figure.

Hariharan et. al. suggested to modify the transmission function, which describes the propagation of light inside an aperture with a radius $a$, by adding a transparent glass ring of optical thickness $\lambda/2$ to the standard pinhole. This both improves the beam flatness of a simple pinhole and flattens the intensity function. In this case we don't have a uniform transmission function, typical of the conventional circular pinholes:

$$t(r) = \begin{cases} 1, & r < a \\ 0, & r \geq a \end{cases} \quad (2)$$

This, inserted inside the U(P) formula, gives in the standard case, a constant. It is not instead in the case proposed by Hariharan. The transmission function is changed into:

$$t(r) = \begin{cases} 1, & r \leq r_m \\ -1, & r > r_m \\ 0, & r \geq a \end{cases} \quad (3)$$

Here, the second ring of radius $r_m$ has to act as a $\lambda/2$ wave plate, that leads to a phase shift of $\pi$, and a t(r) as in Eq. (3). The thickness h of a half-wave plate is defined by:

$$h = \frac{\lambda}{2(n-1)} \quad (4)$$

where n is the index of refraction of the material. In our case, the added layer needs to be 266 nm thick, based on the refractive index of the $SiO_2$ and the wavelength of our setup of 266 nm.

We used the Ansys Lumerical 2021 finite-difference time-domain (FDTD) solver to simulate both standard and HPH pinholes. The incident beam was assumed to have a Gaussian field profile with a mode field diameter of 4 µm. The standard pinholes have a diameter of 2.1 µm. The HPH pinholes were simulated using a Cr hole with diameter of 3.9 µm and a diameter ratio of 85 % for the underlying $SiO_2$ hole. The near to far field projection function was used to calculate the far field after exiting the $SiO_2$ within a distance of 1 meter. The resulting far field distribution of the circular pinholes and the HPH are shown in Fig. 2. The result, normalized with respect to the intensity of the circular pinholes (Fig. 2a) shows that the HPH intensity is reduced to ~ 40 %. When both are normalized and with a focus on the area relevant for our LIL exposures (Fig. 2b), the intensity variation on that area is only of 11%.

### 4. Fabrication process

The fabrication process flow for HPHs demonstrated in this work is shown in Fig. 3a. First, a 6-inch Fused Silica wafer of 1 mm thickness was covered with a 250 nm layer of Cr by evaporation. Then, it was diced into 2.5 cm x 2.5 cm samples.

A positive tone resist was used to structure holes first in Cr and subsequently in $SiO_2$ by electron beam lithography (EBL) and successive etching. The design included variations of the outer hole radius in Cr (and correspondingly the inner one) from 3.6 µm to 2.8 µm, in steps of 0.2 µm. The $SiO_2$ hole diameter percentage ranged between 70 % and 90 % of that of the Cr hole, varied in steps of 5 %. The total design of experiment consists of 25 different pinholes, in line with the simulations and Ref [10]. These design parameters were chosen in accordance with the physical dimensions of our LIL setup with its 266 nm laser wavelength.

The positive tone EBL resist PMMA-950k was deposited through spin coating for a thickness of 900 nm. Each of the pinhole sets was exposed with a Raith EBPG 5200 e-beam system operated at 100 kV acceleration voltage using doses ranging from 253 to 405 µC/cm$^2$. The samples were developed in a 7:3 mix of isopropanol (IPA) and de-ionized water (DI) for 50 sec, then rinsed in IPA and dried using $N_2$ blow dry. Subsequently, the holes were inspected with an optical microscope. Next, wet etching in cerium ammonium nitrate-based chemistry was used to transfer the resist features into the Cr layer. The samples were cleaned in acetone, IPA and dried, and then inspected by scanning electron microscopy (SEM).

For the second EBL, we again used PMMA 950k using the same spin coating process described above and varied the doses from 253 to 405 µC/cm$^2$. Development followed the same recipe detailed above.

Resist features were transferred into the SiO$_2$ substrate by dry etching in an Oxford PlasmaLab 100 tool using a gas mix of CHF$_3$ and Ar. Since the inner hole acts as a λ/2 plate, the target etch depth was 266 nm (Eq. (4)). Finally, the resist was removed from the samples, and they were inspected again by SEM to establish the actual dimensions of the fabricated pinholes. An SEM micrograph of a fabricated HPH is presented in Fig. 3b. Note that the LIL setup requires two identical pinholes to be operated most efficiently.

## 5. Application of Hariharan pinholes in laser interference lithography

We implemented the fabricated HPHs in the LIL setup discussed in Section 2 as an experimental proof-of-concept, choosing gratings with a period of 500 nm as the target. We exposed the gratings on silicon wafers with both conventional pinholes with a diameter of 2.1 µm and a set of two nominally identical HPHs. From the HPHs included in the design parameter variation described above, we selected those with a Cr hole diameter of 3.9 µm and a diameter ratio of 85 % for the SiO$_2$ hole. This choice provided a good compromise between the maximum exposure intensity at the center of the illumination (larger for larger pinholes) and the FWHM of the beam profile (larger for smaller pinholes). As shown in the simulations, the FTP characteristic of HPH comes at the cost of reduced maximum illumination intensity as the total beam energy is spread over a larger area. We then chose the larger HPH diameter, because the center intensity for both pinholes types had to be comparable to ensure comparable exposure times.

SEMI standard 6 inch Si wafers of 675 µm thickness were used as substrates. We spin coated a positive DUV resist for a total resist thickness of 300 nm. In both cases, the first alignment of the setup was to focus the beam spot on the pinholes. The HPHs with their larger diameters received a lower energy, keeping the intensity on the chuck the same as in the case of the standard pinholes. This was verified qualitatively with a comparison of the beam spot images, where the same grey scale variation can be noticed with both pinhole types (Fig. 4 a-b).

During exposure, we used three photodetectors located at the edges of the chuck to monitor quantitatively the intensity. The resulting voltage signal over time is proportional to the dose applied on the resist during the exposure and was used to ensure identical exposure conditions. After exposure, they were first baked for 1 min and 30 s at 140°C and then developed in Megaposit™ MF26 A for 30 s, followed by rinsing in DI water and dried with $N_2$.

The resist features were inspected with SEM, with respect to LW and period. Then the patterns have been transferred into the Si wafers using dry etching (Oxford PlasmaLab 100), with an $SF_6$ based chemistry. The target etch depth has been set to 100 nm, as this depth already provides good contrast during subsequent SEM inspections of the line gratings. Cleaning the wafers with oxygen plasma was the final fabrication step.

## 6. Results and discussion

The LW and the duty cycle (DC) of the features on the fabricated wafers were used as the main benchmark figures. The results of our exposed and etched samples were analyzed taking a series of SEM micrographs. In Fig. 4c and d we show examples of a grating fabricated with standard pinholes and with HPH, respectively. Seven horizontal rows of images were acquired for each wafer, starting at the center of the 6-inch wafer. Subsequent rows of images were arranged in parallel and with a 2 cm distance to each other. In the same row, images were taken with a 5 mm distance, leading to a total amount 150 images per wafer. This required an automated procedure for both acquisition and analysis of the SEM micrographs. We used an in-house software tool that creates scripts which the SEM can execute. The tool allows selecting a series of key parameters, like the magnification and the acceleration voltage, and allows for automated re-focusing at user-defined intervals. The image analysis was automated using the ProSEM software (GenISys GmbH, [18]), which extracts information from SEM images by pattern recognition. We defined the features to be analyzed in the software in areas called region of interest (ROI, see an example in Fig. 5a). Inside each ROI, periodic features are recognized. In our case, the period and LW of the line gratings were extracted and used to calculate statistical data like standard deviations. The software also automatically finds similar features in subsequent images. Thus, by defining the ROI accordingly, all lines can be measured at multiple positions, reducing measurement errors.

The LW variation across the wafer is a good measure for assessing the benefits of our HPHs. As explained in Section "theory", they provide a higher uniformity in terms of FWHM, which should translate directly to a lower variation in LW.

The distribution of LWs across both wafers exposed with a conventional and HPHs is shown in Fig. 5 b-c. Each point in the plots corresponds to one image acquisition and its average LW based on the automated analysis. The points missing in Fig. 5 b-c correspond to failures of the autofocus during the image acquisition, which prevented their analysis. The plots show a much larger LW variation for the wafer exposed with conventional pinholes (Fig. 5b) compared to the one fabricated using HPHs (Fig. 5c).

We have then focused on the wafer's center line, since the interference pattern is created with a periodicity along the x direction. The LW variation along this center line for both wafers is shown in Fig. 5d. The experimental data reflect the simulations, with a flat part extending over a larger area for HPHs than for conventional pinholes. Quantitatively, the HPHs show a LW variation of 27.5 % while the conventional pinholes show 47.2 % over the 6-inch wafer.

## 7. Conclusions

We have experimentally demonstrated a method to account for errors originating in the Gaussian profile of spatially filtered laser beams. Based on a concept originally theorized by Hariharan et al. [11], we have developed a fabrication process flow for this new class of optical elements, i.e. Hariharan pinholes. We have implemented HPHs in a laser interference lithography system to demonstrate their functionality and assess their performance in an application scenario as a proof of concept. The setup used HPHs as part of its spatial filters to fabricate line gratings with a nominal period of 500 nm and a target duty cycle of 43 %. Structures fabricated by LIL using HPHs have a LW variation of 27.5 %, compared to 47.2% when using standard pinholes. This clearly demonstrates that HPHs are a valuable tool to homogenize beam shapes in specific laser applications.

**Acknowledgements**: This work has received funding from the European Union's Horizon 2020 research and innovation program under the Marie Skłodowska-Curie grant agreement No 813159 (GREAT) and the German Federal Ministry for Economic Affairs and Climate Action under grant agreement 49MF210208 (PESOS).

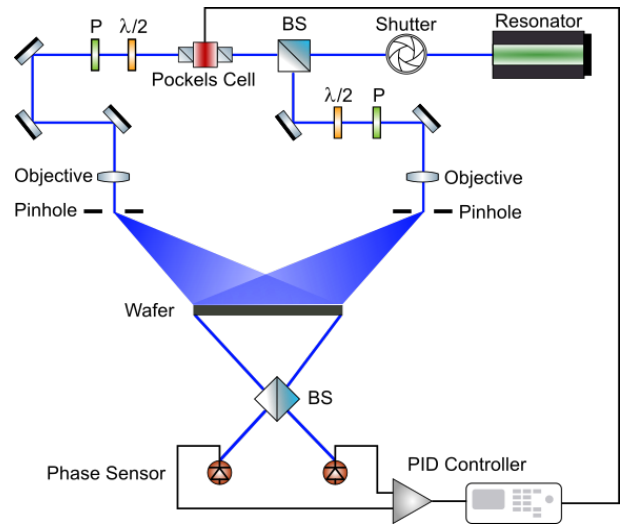

Fig. 1: Scheme of a laser interference lithography setup: after the resonator and a shutter, the beam is divided by a beam splitter (BS); each beam path then contains a half-wave plate (λ/2), a polarizer (P) (one of the two also a Pockels cell) and mirrors that leads the two beams to spacial filters, consisting of an objective and a pinhole, before interfering on a substrate.

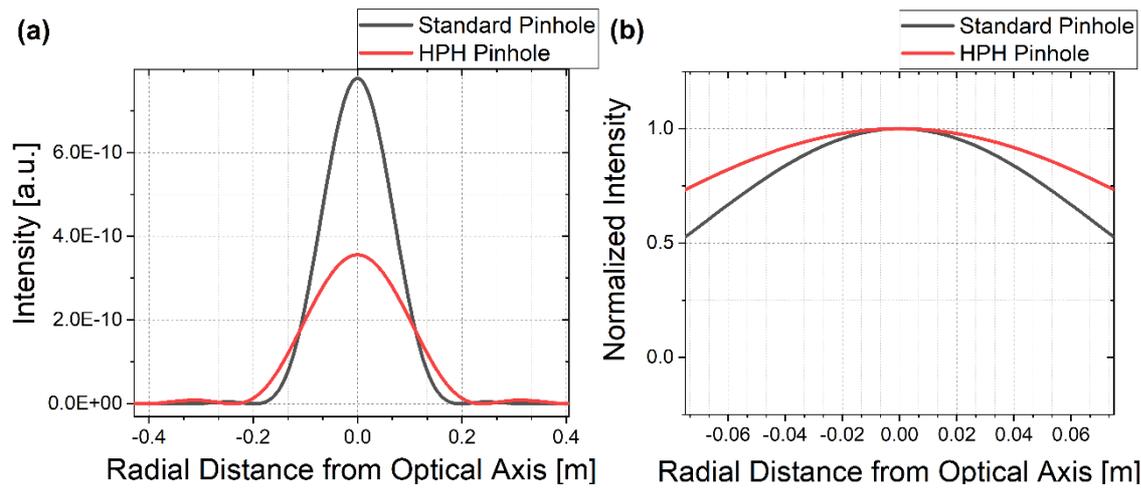

Fig. 2: Intensity profile calculated (Ansys Lumerical 2021 finite-difference time-domain FDTD solver) from the Fraunhofer diffraction model for a conventional d = 2.1 µm (black) and a d = 3.93 µm Hariharan pinhole (red), (a) both normalized to the intensity in the case using a standard pinhole (b) both normalized to 1, with a focus on the functions for an area of 15 cm from the axis of the beam.

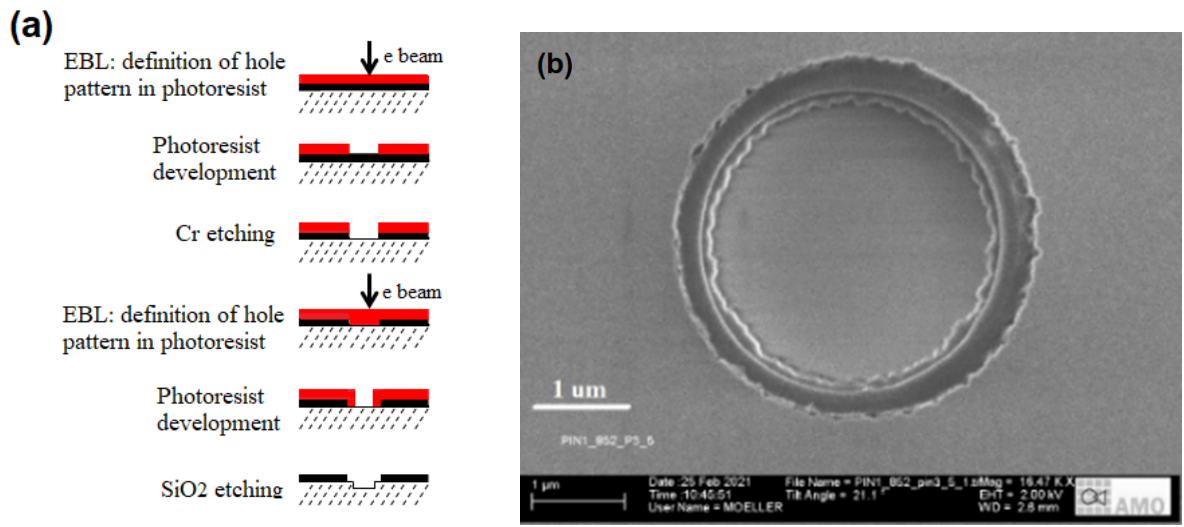

Fig. 3: (a) Scheme of the HPH fabrication steps; (b) SEM image of a fabricated HPH. The outer ring is a hole in Cr and has a diameter of 3.9 µm, while the inner one in SiO2 and has a diameter of 85% of the one in Cr.

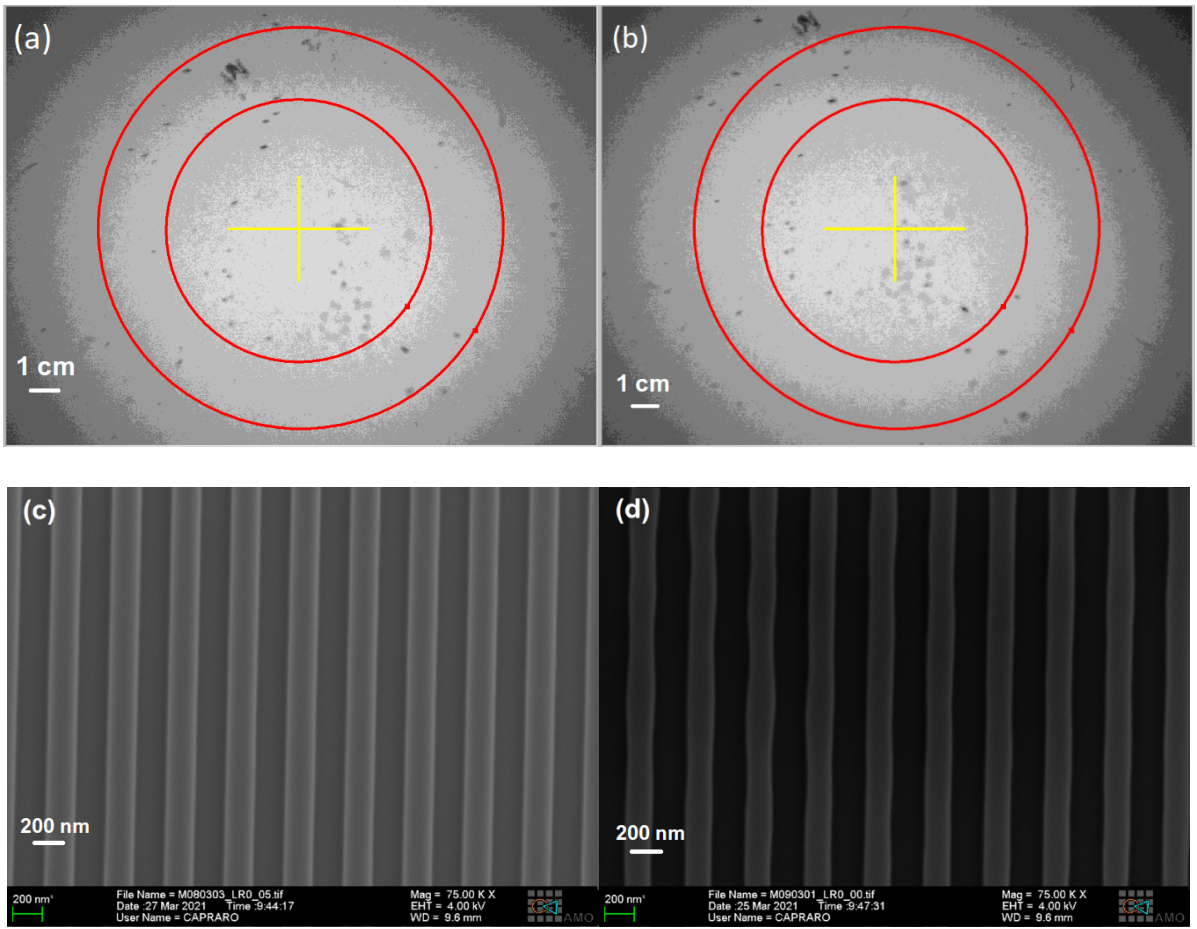

Fig. 4: Beam spot on the chuck, produced with (a) a standard pinhole of d = 2.1 µm and (b) a HPH of 3.9 µm. The images were acquired by a photodiode positioned in front of the chuck. The yellow cross is the center of the chuck while the two red circles highlight the two brightest shades of grey, indicating a high beam intensity. Following, SEM micrograph of a grating fabricated with LIL, using a set of standard pinholes (c) and with the HPH (d).

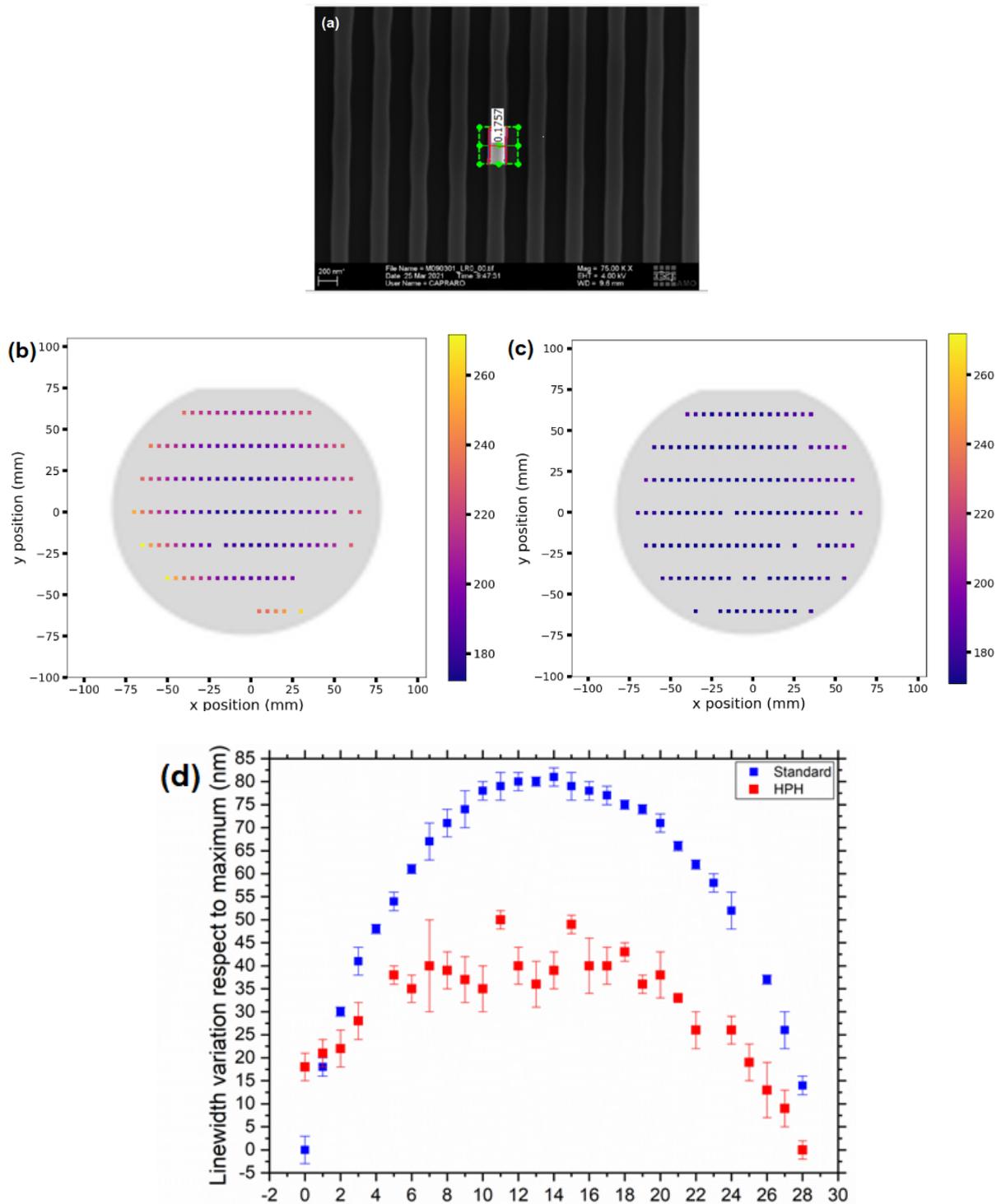

Fig. 5: (a) Screenshot of the analysis of a SEM micrograph made with ProSEM. Measured LW variation on wafers exposed with (a) standard pinholes and (b) with HPHs. The rows have a distance of 2 cm, while the points in the same row are placed on a 5 mm grid. (d) LW variation across the central diameter of line gratings fabricated with standard pinholes (blue) and HPH (red).

**Supporting Information for "Phase mask pinholes as spatial filters for laser interference lithography"**

G. Capraro,[1,2] M. Lipkin,[1] M. Möller,[1] J. Bolten[1], M. C. Lemme[1,2,*]

[1] Chair of Electronic Devices, Faculty of Electrical Engineering and Information Technology,

RWTH Aachen University, Otto-Blumenthal-Straße 2, 52074 Aachen

[2] AMO GmbH, Otto-Blumenthal-Straße 25, 52074 Aachen

*lemme@amo.de

**1. Derivation of Airy formula**

A wave that impinges with normal incidence on a circular pinhole of radius a, is diffracted by it. In the far field area, where source as long as point of observation P are much more distant from the aperture than its diameter d, the complex amplitude U of the diffracted wave, in the point P of observation and written in polar coordinates is (Fraunhofer diffraction) [1]:

$$U(P) = 2\pi C \int_0^a J_0(k\rho w)\rho d\rho \quad (1)$$

where $\rho$ is one of the polar coordinates of a point in the aperture; w is one of the coordinates of the point P in the diffraction pattern; $J_0$ is the Bessel function of order $0^{th}$, defined as

$$J_0(x) = \frac{1}{2\pi}\int_0^{2\pi} e^{ix\cos(\alpha)} d\alpha \quad (2)$$

In a more general description of the U(P) expression, the transmission function is also present. This function describes how the aperture transmits a wave propagating through it. In the case of a circular pinhole, this transmission function is just a constant. The constant C instead contains information about the wavelength and power of the beam:

$$C = \frac{1}{\lambda R}\left(\frac{P}{D}\right)^{\frac{1}{2}} \quad (3)$$

where λ is the beam wavelength; R is the distance from the origin in the plane containing the aperture to the point at which the axis passing through the source intersects the plane of observation; $P$ is the total power incident upon the aperture; $D = \pi a^2$ is the area of the opening.

Using the properties of the Bessel functions, Eq. (1) can be also written as

$$U(P) = CD \frac{2J_1(kaw)}{kaw} \qquad (4)$$

The intensity instead is given by

$$I(P) = |U(P)|^2 = I_0 \left[\frac{2J_1(kaw)}{kaw}\right]^2 \qquad (5)$$

that is the well-known Airy function.

Each beam that is going to interfere will have the first intensity minimum, taking into account this simplified description in Eq. (5), where $J_1(kaw) = 0$.

## 2. Comment on increased uncertainty for the LW calculations of exposures with HPH

Looking at Fig. 4: Beam spot on the chuck, produced with (a) a standard pinhole of d = 2.1 μm and (b) a HPH of 3.9 μm. The images were acquired by a photodiode positioned in front of the chuck. The yellow cross is the center of the chuck while the two red circles highlight the two brightest shades of grey, indicating a high beam intensity. Following, SEM micrograph of a grating fabricated with LIL, using a set of standard pinholes (c) and with the HPH (d). it is noteworthy that the LW measurements of the HPH's wafers show an increase measurement uncertainty compared to the wafer exposed using conventional pinholes. This can be attributed to the fact that the FTP effect of the HPHs can lead to increased scattering of light on the surface of the optical table used for the LIL setup. As a result, a LW fluctuation along the y axis of the exposure is visible in some of the SEM micrographs obtained from the wafer exposed using HPHs (see Fig. 4c and Fig. 4d, as a comparison of a grating using standard pinholes and HPH). One easily implementable option to counter this effect is to increase the height in which the beam path is led over the optical table or further improve the size of the pinholes and adapt them properly to the size of the exposure field.

**3. LW variation over a set of wafers fabricated with HPH**

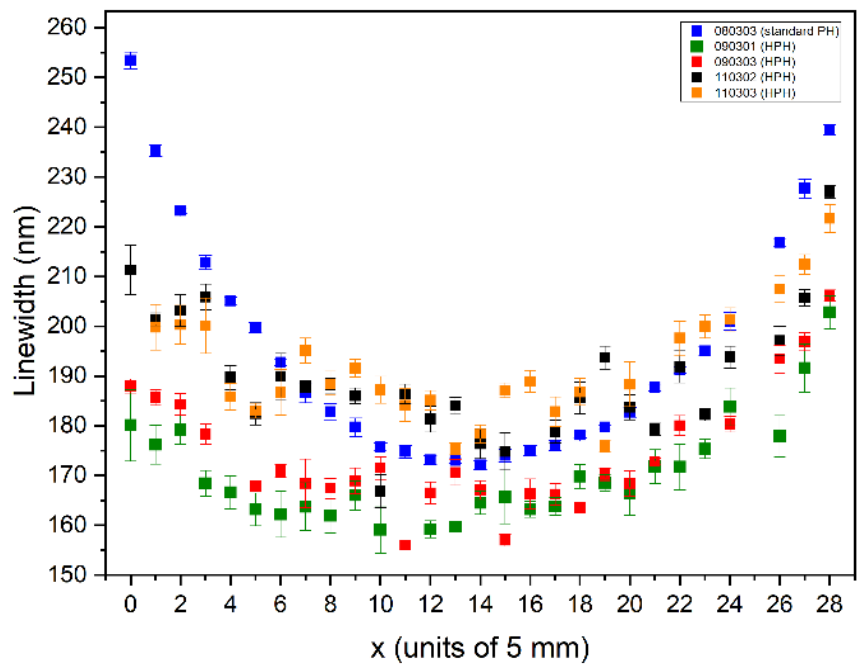

Fig. SI 1: LW variation across the central diameter of GWS fabricated with standard pinholes (blue) and HPH (red, green, orange and black).